\documentstyle[amstex,aps,prl,multicol,epsfig,amssymb]{revtex}
\begin{document}
%\author{V. A. Ivanov, J. J. Betouras, and F. M. Peeters \\
%Departement Natuurkunde, Universiteit Antwerpen (UIA),\\
%Universiteitsplein 1, B-2610 Antwerpen, Belgium}
%\title{MgB$_{2}$: superconductivity and pressure effects}
%\maketitle

{\center  {\Large MgB$_{2}$: superconductivity and pressure effects \\[1.00cm]}
{\large {V. A. Ivanov, J. J. Betouras, and F. M. Peeters \\
Departement Natuurkunde, Universiteit Antwerpen (UIA),\\
Universiteitsplein 1, B-2610 Antwerpen, Belgium\\[1.50cm]  } } } 

\begin{abstract}
We present a Ginzburg-Landau theory for a two-band superconductor
with emphasis on MgB$_{2}$.
Experiments are proposed which lead to identification of the possible
scenarios: whether both $\sigma $- and $\pi $-bands superconduct or $\sigma $%
-alone. According to the second scenario a microscopic theory of
superconducting MgB$_{2}$ is proposed based on the strongly interacting $%
\sigma $-electrons and non-correlated $\pi $-electrons of boron ions. The
kinematic and Coulomb interactions of $\sigma $%
-electrons provide the superconducting state with an anisotropic gap of $s$%
*-wave symmetry. The critical temperature $T_{c}$ has a non-monotonic
dependence on the distance $r$ between the centers of gravity of $\sigma $-
and $\pi $-bands. The position of MgB$_{2}$ on a bell-shaped curve $T_{c}$ ($%
r)$ is identified in the overdoped region. The derived superconducting
density of electronic states is in agreement with available experimental
and theoretical data. It is argued that the effects of pressure are crucial
to identify the microscopic origin of superconductivity in MgB$_{2}$.
Possibilities for $T_{c}$ increase are discussed.
%\newline
%\newline
\end{abstract}

The discovery of superconductivity in MgB$_2$ \cite{nagamatsu} 
poses many interesting and fundamental questions regarding the nature
of the superconducting state as well as the possibility of multi-band
superconductivity. 
The crystal belongs to the space group $P6/mmm$ or $AlB_2$-structure 
where borons are packed in honeycomb layers alternating with hexagonal layers 
of magnesium ions.
The ions Mg$^{2+}$ are positioned above the centers of hexagons formed by boron
sites and donate their electrons to the boron planes.
The electronic structure is organized by the narrow energy bands with near
two-fold degenerate $\sigma$-electrons and the wide-band $\pi$-electrons. 
Without any of the
lattice strain, the $\sigma$ dispersion relations are slightly splitted 
due to the two boron atoms per unit cell. The corresponding portions of the
Fermi surface consist of coaxial cylinders along the $\Gamma$ - 
A symmetry direction of the Brillouin zone (BZ), whereas the $\pi$- bands are strongly dispersive.
In the following sections we present a Ginzburg-Landau (GL) analysis of the
two-band superconductor with emphasis on MgB$_2$, we apply it in
pressure experiments which potentially distinguish the different
superconducting bands \cite{betouras1}, then we provide a microscopic model. 

\section{Ginzburg-Landau analysis and pressure effects}

We distinguish different possibilities:

{\it Case I: Both $\sigma$ and $\pi$ bands superconduct.}
The GL free energy functional for MgB$_2$ in this case can be written as :
\begin{eqnarray}
\nonumber
F = \int d^3r \{ \frac{1}{2m_{\sigma}} |\vec{\Pi} \psi_{\sigma}|^2 + 
\alpha_{\sigma} |\psi_{\sigma}|^2 + \beta_{\sigma} |\psi_{\sigma}|^4 + 
\frac{1}{2m_{\pi}} 
|\vec{\Pi} \psi_{\pi}|^2 + \alpha_{\pi} |\psi_{\pi}|^2 + 
\beta_{\pi} |\psi_{\pi}|^4 \\
+ r (\psi_{\sigma}^* \psi_{\pi} + \psi_{\sigma} \psi_{\pi}^*) 
+ \beta (|\psi_{\sigma}|^2 |\psi_{\pi}|^2)\ +
\frac{(\vec{\nabla} \times \vec{A})^2}{8 \pi} \},
\end{eqnarray}
where $\vec{\Pi} = -i \hbar \vec{\nabla} - 2e/c \vec{A}$ and $\vec{A}$ 
is the vector potential, $\alpha_{\sigma,\pi} = \alpha_{\sigma,\pi}^0 
(T-T_{c\sigma,\pi}^0)$.

(i) If r = 0 then Eq. 1 is the free energy for two bands 
without pairing transfer (Josephson coupling) between them. 
The quartic term which
mixes the two order parameters reflects the fact
that although the two bands are different, there is a constraint coming
from the common chemical potential which connects them in equilibrium.
It is also the realization and a measure of the strength 
of the interband interaction of quasiparticles and it 
affects the critical temperatures.
The absence of a  bilinear in the two order parameters term facilitates the
observation of the Leggett's mode in the Josephson tunneling between 
a two-band superconductor and an ordinary superconductor \cite{agterberg}.
The onset of the superconducting state in one band does not imply the onset
in the other.
To investigate this possibility in MgB$_2$ an experiment which checks
the splitting of the 
transitions due to the lattice deformation by the strain fields can be
decisive.
In the MgB$_{2}$ the compression due to pressure 
is anisotropic\cite{goncharov,tissen,vogt}. 
According to Ref.~%
\cite{goncharov} the compressibility along the $c$-axis is almost twice
larger that of the plane compressibility.
Therefore application of uniaxial pressure on single crystals 
will affect differently the two gaps
and the result will be a splitting of the critical temperatures, even if
at ambient pressure they appear to be the same. 
This can be measured through a specific heat experiment under pressure.  
Following Ozaki's formulation
\cite{ozaki,sigrist}, we add to the GL functional Eq.(1) the 
term which couples the order parameters, in second 
order,  with the strain tensor $\epsilon$ to first order. 
\begin{equation}
F_{strain} = - C_1(\Gamma_1) [\delta (\epsilon_{xx}+\epsilon_{yy}) +
 \epsilon_{zz} ] |\psi_{\sigma}|^2 
- C_2(\Gamma_1) (\epsilon_{xx}+\epsilon_{yy} +
 \epsilon_{zz}) |\psi_{\pi}|^2 , 
\end{equation}
where $C_{1,2}(\Gamma_1)$ are coupling constants, $\delta$ is given in terms of
the elastic constants of the material $\delta = (c_{11}+c_{12}-c_{13})/(c_{13}
-c_{33})$.
The effect of the pressure on these two order parameters, each one belonging to
an one-dimensional representation  is to shift the critical
temperatures. 
The new critical temperatures can be found,
by solving the coupled equations for
$|\psi_{\sigma,\pi}|^2$ and then 
setting  the coefficients of $|\psi_{\sigma,\pi}|^2$ to zero. 
The result is : $T_{c\sigma,\pi} =  T_{c\sigma,\pi}^0 - \eta_{\sigma,\pi} 
\times p$,
where for uniaxial pressure $\eta_{\pi} > \eta_{\sigma} > 0$. 
In order to detect the splitting, the pressure must be above a certain value
$p_{min}$ which depends on the resolution of the experiment,
so that the difference in the two critical temperatures 
$\Delta T_c = |T_{c\sigma}^0 -T_{c\pi}^0+ \eta_{\sigma}-\eta_{\pi}| 
p_{min}$ is experimentally detectable.

(ii) If r $\neq$ 0 then the pair transfering term is present and it means 
that the onset
of superconductivity in one band implies automatically the appearance of
superconductivity in the other. There is a single observed T$_c$ with a 
 different pressure dependence. Analyzing the equations which result
from the minimization of the GL free energy we get 
 in the regime $T_{c\sigma}^0 > T > T_{c\pi}^0$ :
\begin{equation}
\alpha_{\pi} \alpha_{\sigma} =  r^2
\end{equation}
This gives the pressure dependence of the $T_c$:
\begin{equation}
T_c(p) = \frac{1}{2} \left[ T_{c\sigma}^0 + T_{c\pi}^0 - 
(\eta_{\pi}+\eta_{\sigma})p \right] + 
\frac{1}{2} 
\left[(T_{c\sigma}^0 - T_{c\pi}^0 + (\eta_{\pi} - \eta_{\sigma})p)^2 + 
a^2\right]^{1/2}
\end{equation}
and $a^2 = 4r^2/
(\alpha_{\pi}^0 \alpha_{\sigma}^0)$. Deviations from a straight line at 
moderate values of pressure can be attributed to the two bands.
 
{\it Case II: Only the $\sigma$- bands superconduct.}
In this case there is one order parameter initially
and the GL free energy functional is the usual one.
The dominant physical situation is an important change in the 
electronic properties of the material under pressure, 
because the band ($\sigma_2$) which is below the Fermi
level at ambient pressure \cite{an} can, partially, overcome the barrier 
and get above the
Fermi level at a certain value of pressure (crossover pressure), 
restoring the degeneracy of the two $\sigma$ bands at point $\Gamma$. 
Then there will be a crossover from
a superconducting state $\psi_1$ to a state $\psi_1 + \psi_2$. 
To understand this effect we need to write the GL
free energy taking into account the second order parameter of the same
symmetry as well as the hexagonal symmetry of the 
boron layers. 
The irreducible representation of the $D_{6h}$ group are four one-dimensional 
ones (three of them have line nodes) and two two-dimensional ones 
\cite{sigrist}. There are experimental data compatible with {\it s}-wave 
or an anisotropic $s^*$-wave order parameter.
and a penetration depth experiment \cite{panagopoulos}
which suggests nodes or anisotropic $s^*$-wave. 
Therefore we consider the gap to be  
a  function of $k_x$ and $k_y$ with a smooth modulation in the $k_z$ direction
as in Ref. \cite{maki}.
As a consequence the most promising candidates for the initial order 
parameter 
are the basis functions of the $\Gamma_1$ representation of the 
$D_{6h}$ group. 
To construct the functional of the free energy for two order parameters,
we need to take
into account the decomposition of the terms containing the derivatives
$\vec{\Pi}$, which belong to the $\Gamma_5$
representation. Then the decomposition 
\begin{equation}
{\Gamma_5}^{*} \otimes {\Gamma_1}^{*} \otimes 
{\Gamma_5} \otimes \Gamma_1  = \Gamma_1 + \Gamma_2 ,
\end{equation} 
contains the $\Gamma_1$ representation once and suggests  that there is 
precisely one term quadratic in ${\Pi}_{x,y}$ which mixes the two order 
parameters (to first order each). There is also a quadratic term which mixes 
the two components, according to the  trivial decomposition $\Gamma_1 \otimes
\Gamma_1$ and a quartic term as well. 
Taking into accounts all the usual terms, we obtain the following
expression for the free energy:
\begin{eqnarray}
\nonumber
F = \int d^3r \{ \frac{1}{2m_1} |\vec{\Pi} \psi_1|^2 + 
\alpha_1 |\psi_1|^2 + \beta_1 |\psi_1|^4 + \frac{1}{2m_2} 
|\vec{\Pi} \psi_2|^2 + \alpha_2 |\psi_2|^2 + \beta_2 |\psi_2|^4 +\\
\gamma_1 (\psi_1^* \psi_2 + c.c.) 
+ \gamma_2 (\Pi_x \psi_1 \Pi_x^* \psi_2^*
+ \Pi_y \psi_1 \Pi_y^* \psi_2^* + c.c.) + \beta_{3} (|\psi_1|^2 |\psi_2|^2)\
+ \frac{(\vec{\nabla} \times \vec{A})^2}{8 \pi} \}.
\end{eqnarray}
We have used the fact that the $\gamma_1$-term favors the coupling of linear
combination of the two order parameters with phase difference 0 or $\pi$ 
between them \cite{betouras},
therefore a term $\psi_1^{2*}\psi_2^2 + c.c.$ is incorporated into the 
$\beta_3$-term of the free energy. 
The resemblence between Eq.(1) and Eq.(7) is apparent; in the first case
the mixing terms are a consequence of the interactions between the two bands,
in the second case it is due to the modification of the existing
gap in the electronic spectrum. 

With the present approach we are in the position to describe the crossover 
and to consider the possibility of different 
effective masses and parameters (the inclusion of the effect at $p_c$).  
The effect of pressure can be taken into account as usual \cite{bob} 
with the modification due to the particular physical
situation by writing $\alpha_2=\alpha_2^0 (T-T_c(p)) \Theta(p-p_c)$, 
$\gamma_1= \gamma_1^0 (p-1) \Theta(p-p_c)$. 
The additional term which couples the order parameter  with the strain 
tensor $\epsilon$, in the regime where $p > p_c$ is : 
\begin{equation}
\nonumber
F_{strain} = - C(\Gamma_1) [(1+\delta) (\epsilon_{xx}+\epsilon_{yy}) +
2 \epsilon_{zz} ] \times 
|\psi_1 + \psi_2|^2 . 
\end{equation}
The coupling to strain affects both order parameters 
in the same way.  

The approximate value of the  crossover pressure 
can be estimated as follows. 
The energy difference  between the two subbands is approximated by $%
\left( 1-n_{\sigma }\right) ^{2}\Delta $, where the fraction of
superconducting electrons is $1-n_{\sigma }\sim 0.03$, {\it i.e.} the carrier
density per boron atom in Mg$B_{2}$. The approximate value of the
crossover pressure $p_{c}$ which suppresses the deformation potential 
$\Delta $\
can be estimated by the expression 
%\[
$p_{c}\Omega \sim \left( 1-n_{\sigma }\right) \sqrt{\Delta }$
%\text{,} \]
where $\Omega =30$\AA $^{3}$ is a unit cell volume of Mg$B_{2}$ and $%
\Delta =0.04eV^{2}$ is the deformation potential for a boron displacement $%
u\sim 0.03$\AA\ \cite{yildirim}. Using these parameters, we get a
crossover pressure of $p_c \sim 30 GPa$. This estimation shows, 
that an applied pressure at a realistic value
influences drastically the electronic structure, producing the
degeneracy of two bands which are initially splitted.
Superconductivity in the second band will occur at lower 
values of the estimated $p_c$. The requested band overlapping 
around the Fermi energy will also occur at lower values due to the 
corrugation of the Fermi surfaces and the already existing strains in the
material and the anisotropic compressibility. These considerations 
make the above estimation an upper limit of $p_c$.
The degree to which shear stresses of the sample and the
surrounding fluid under pressure affect the data of
pressure measurements is still under investigation \cite{schilling}. 
The physics of this crossover is similar in spirit to an electronic topological
transition of 5/2 kind \cite{lifshits}. In Fig.1 in a model calculation, 
we illustrate
the expected behavior of T$_c$ and  
the form of the order parameter as a function of pressure
at fixed temperature and schematically the kink at p$_c$ of T$_c$. 
The anomaly close to $p_c$ can be detected directly in 
a penetration depth experiment under pressure.
%\begin{minipage}[b]{.99\linewidth}
\begin{figure}
\begin{center}
\epsfig{figure=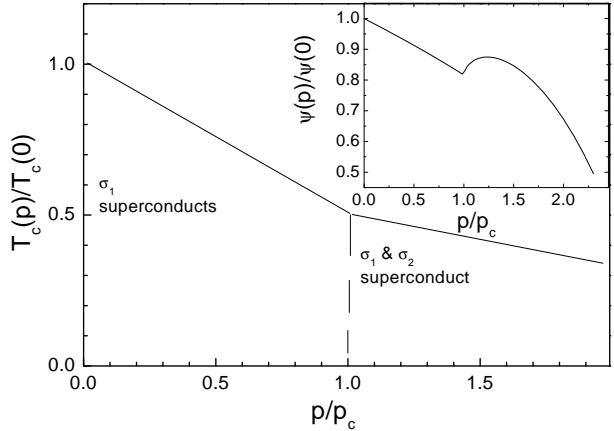,height=6.7cm}
\end{center}
\caption{The form of T$_c$ as a function of pressure for case II,
$p_c$ is the crossover pressure where the second band becomes
superconducting as well.  
Inset : The normalized absolute value of the order parameter 
$\psi(p)/\psi(0) = |\psi_1(p) + \psi_2(p)|/|\psi_1(0)|$ at 
$T= 0.7 T_{c1}$. 
The chosen parameters are: $\alpha_1 = 2 
(T/T_{c1}-1-0.1 p/p_c)$, $\alpha_2 = (T/T_{c1}-0.8-0.1 p/p_c)$, 
$\beta_1= \beta_2 =1$, $\beta_3 = \Theta(p/p_c-1)$, 
$\gamma_1= 0.4 (p/p_c-1) \Theta(p/p_c-1)$.}
\label{fig:fig2}
\end{figure}
%\end{minipage}

The picture we provide for the second case is complimentary to the first.
Both cases may be realized since the effects can be detected at
different values of pressures. In case I, the value of pressure 
necessary to see the effects is limited either by the experimental 
resolution or by the validity of the Eq. 4  at low pressure.
The specific symmetry
of the order parameters is only used to produce the additional terms
due to strain. In case II there is a larger predicted value for
the crossover pressure and there are 
more specific assumptions regarding the symmetry of the order parameter.
Early pressure experiments \cite{monteverde} demonstrated the overall 
decrease of $T_c$ with pressure which was attributed
to the loss of holes. In two of the samples there is a linear dependence
of $T_c$ on pressure and in two others a weak quadratic dependence.
We stress that the samples are polycrystalline and the experiment is 
effectively under hydrostatic pressure. Also the degree of nonstoichiometry 
was not known. The almost linear dependence for a wide range of 
pressures, makes the GL functional as presented, valid for the MgB$_2$.
Experiments on single crystals
will be able to verify the effects which are described.
We do not attempt at the moment any fitting of experimental data
because there is no experimental consensus on the different values of 
key parameters of the theory (e.g. there is a wide range of published data
on the value of $dT_c/dp$ \cite{schilling}). 

\section{Microscopic model}
The presented microscopic model of superconductivity in MgB$%
_{2}$, is based on the correlated $\sigma $- and the non-correlated $\pi $%
-band. The negatively charged boron layers and the positively charged
magnesium layers provide a lowering of the $\pi $-band with respect to the $%
\sigma $-band which was also noticed first in band calculations \cite{3,3b}.
In this microscopic treatment the $\sigma$-bands are treated as degenarate.
%Tight-binding estimates show that the bonding $\sigma $-band, close to the
%Fermi level, can be taken as 
%doubly degenerate (it has $E$-symmetry \cite{4,5}). 
The quasi-2D $\sigma $-electrons are more localized than the 3D $\pi $%
-electrons. This leads to an enhancement of the on-site electron
correlations in the system of degenerated $\sigma $-electrons. They are
taken to be infinite, while the intersite Coulomb
interactions and electron-phonon interactions simply shift the on-site
electron energies. After carrying out a fermion mapping to $X$-operators the
Hamiltonian becomes: 
\begin{eqnarray}
\nonumber
H&=&\sum\limits_{p,s}t^{(\sigma
)}(p)X^{so}(p)X^{os}(p)+V\sum\limits_{\left\langle i,j\right\rangle
}n_{\sigma }\left( i\right) n_{\sigma }\left( j\right)
-r\sum\limits_{i}n_{\pi }\left( i\right) +  \\
&+&\sum\limits_{p,s}\varepsilon _{0}^{(\pi )}(p)\left[ \pi _{s}^{+}(p)\pi
_{s}(p)+H.c.\right] -\mu \sum\limits_{i}\left[ n_{\sigma }\left( i\right)
+n_{\pi }\left( i\right) \right] .  
\end{eqnarray}
Here, the $X$-operators
describe the on-site transitions of correlated $\sigma $-electrons between
the one-particle ground states (with spin projection $s=\pm $) and the empty
polar electronic states ($0$) of the boron sites and $\mu $ is the chemical
potential. The wide $\pi $-band is shifted with respect to the $\sigma $%
-band by an energy $r$, comprising the mean-field $\pi $%
-electron interactions and the electron-phonon interactions. 
We included in Eq. $\left( 7\right) $\ also the
nearest neighbour Coulomb repulsion $V$\ of $\sigma $-electrons, which is
essential for the low density of hole-carriers in MgB$_{2}$. The
mutual hopping between $\sigma $- and $\pi $-electrons is assumed to be
negligible due to the characteristic space symmetry of the orbitals
involved. A schematic view of the model is presented in Fig. 2. 
\begin{figure}
\begin{center}
\epsfig{figure=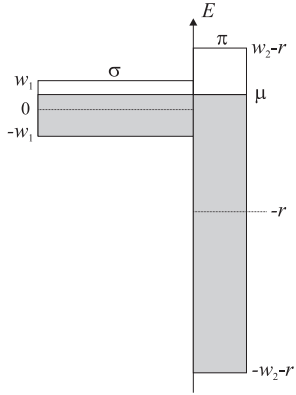,height=5.2cm}
\end{center}
\caption{A schematic view of the energy band diagram. The degenerate $
\sigma $-electrons are represented by a lower correlated band.}
\label{fig:nato2}
\end{figure}

The diagonal 
$X$-operators ($X^{\cdot }\equiv X^{\cdot \cdot }$) satisfy the
completeness relation, $X^{0}+4X^{s}=1$, and their thermodynamic averages ($%
\left\langle X^{0}\right\rangle $, $\left\langle X^{s}\right\rangle $) are
the Boltzmann populations of the energy levels of the unperturbed on-site
Hamiltonian in $\left( 7\right) $. Due to the orbital and spin degeneracy
the $\sigma $-electrons occupy their one-particle ground state with density $%
n_{\sigma }=4\left\langle X^{s}\right\rangle $ per boron site. The
correlation factor for the degenerated $\sigma $-electrons is 
$f=\left\langle X^{0}+X^{s}\right\rangle =1-3\left\langle X^{s}\right\rangle
=1-\frac{3}{4}n_{\sigma }$.
It plays an important role in all numerics starting from their
unperturbed (zeroth order) Green's function, $D_{\sigma }^{\left( 0\right)
}\left( \omega \right) =f/\left( -i\omega _{n}-\mu \right) $, whereas for $%
\pi $-electrons $D_{\pi }^{\left( 0\right) }\left( \omega \right) =1/\left(
-i\omega _{n}-\mu -r\right) $. Note that for the conventional Hubbard model
the correlation factor in the paramagnetic phase is $1-n/2$ (see \cite{7}
and Refs. therein). 

The energy dispersion of both bands, $\xi \left(
p\right) =ft\left( p\right) -\mu $ and $\varepsilon \left( p\right)
=\varepsilon _{0}\left( p\right) -r-\mu $, are governed by the zeros of the
inverse Green's function 
$D^{-1}\left( \omega ,p\right) =\text{diag}\left\{ D_{\sigma }^{-1}\left(
\omega ,p\right) ;D_{\pi }^{-1}\left( \omega ,p\right) \right\} = 
\text{diag}\left\{ \frac{-i\omega _{n}+\xi \left( p\right) }{f};-i\omega
_{n}+\varepsilon \left( p\right) \right\}$,
which follows from the Dyson equation 
$D^{-1}\left( \omega ,p\right) =D^{\left( 0\right) -1}\left( \omega \right) +%
\widehat{t}\left( p\right) \text{,} $
to first order with respect to the tunneling matrix $\widehat{t}\left(
p\right) =$diag$\left\{ t\left( p\right) ;\varepsilon _{0}\left( p\right)
\right\}$. 
%$where diag$\left\{ A;B\right\} =\left( 
%\begin{array}{cc}
%A & 0 \\ 
%0 & B 
%\end{array}
%\right) $.

The chemical potential $\mu$ for the MgB$_{2}=$Mg$^{++}$B$^{-}\left(
p^{2}\right) _{2}$=Mg$^{++}$B$^{-}\left( p^{n_{\sigma}}p^{n_{\pi}}\right)
_{2}$-system (Eq. $\left( 1\right) $) obeys the equation for the total
electron density per boron site 
$n_{\sigma}+n_{\pi}=2$, 
with the partial electron densities: 
\begin{equation}
n_{\sigma,\pi}=2T\sum \limits_{n,p}e^{i\omega\delta}D_{\sigma,\pi}\left(
\omega,p\right) \equiv n_{\sigma,\pi }\left( r,\mu\right) . 
\end{equation}
The electron densities satisfy the requirements 
$0<n_{\sigma}<1$ (due to the correlation factor f,
providing the quarter-fold narrowing of the degenerate $\sigma$-band) and $%
0<n_{\pi}<2$. For any energy difference $r$\ between the\ $\sigma$- and \ $%
\pi$-bands the chemical potential $\mu$\ has to be such that the 
constraint $n_{\sigma}+n_{\pi}=2$ is satisfied.

From the system of Eqs. of the constraint and (9), for a flat density of
electronic states (DOS hereafter) $\rho_{\sigma,\pi}\left( \varepsilon
\right) =\left( 1/2w_{1,2}\right) \theta\left( w_{1,2}^{2}-\varepsilon
^{2}\right) $ with half-bandwidths $w_{1}$ and $w_{2}$ for $\sigma$- and $%
\pi $-electrons, respectively) we
derive the chemical potential of the $A^{2+}B\left(
p^{n_{\sigma}}p^{n_{\pi}}\right) _{2}$ systems as 
\begin{equation}
\mu=\frac{w_{2}-5r}{5w_{1}+4w_{2}}w_{1.} 
\end{equation}
The non-correlated $\pi$-electrons play the role of a reservoir for the $%
\sigma$-electrons. In the energy dispersion $\xi\left( p\right) $\ for the $%
\sigma$-electrons, the correlation factor $f$ can be
expressed also via the parameters $w_{1},w_{2}$ and $r$
as $f=\left( 2w_{1}+w_{2}+3r\right) /\left( 5w_{1}+4w_{2}\right)$.

The anomalous self-energy for the $\sigma$-electrons (Fig. 3) is written
self-consistently as 
\begin{equation}
\overset{\vee}{\Sigma}\left( p\right) =T\sum \limits_{n,q}\Gamma_{0}\left(
p,q\right) \frac{\overset{\vee}{\Sigma}\left( q\right) }{\omega_{n}^{2}+%
\xi^{2}\left( q\right) +\overset{\vee}{\Sigma}\left( q\right) }, 
\end{equation}
where the vertex $\Gamma_{0}$ is determined by the amplitudes of the
kinematic and Coulomb interactions such that $\Gamma_{0}\left( p,q\right)
=-2t_{q}+V\left( p-q\right) $. We do not include the other kinematic
vertices in $\Gamma_{0}$ which are essential at a 
moderate concentration of carriers 
\cite{7}. 
\begin{figure}
\begin{center}
\epsfig{figure=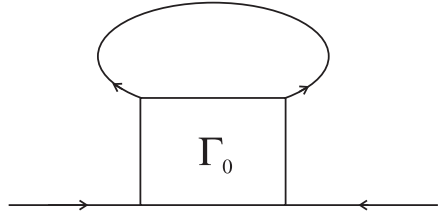,height=2.8cm}
\end{center}
\caption{The anomalous self-energy $\overset{\vee }{\Sigma }\left(
p\right) $ for $\sigma $-electrons. The solid line is an anomalous Green's
function. }
\label{fig:nato3}
\end{figure}

In
momentum space the Coulomb repulsion between the nearest neighbours (Eq. $%
\left( 7\right) $) reflects the tight-binding symmetry of the boron
honeycomb lattice \cite{ssc}. Near the $\Gamma-A$ line of the Brillouin zone
the Coulomb vertex can be factorized as 
$V\left( p-q\right) =2\beta t\left( p\right) t\left( q\right)$,
where the parameter $\beta=V/6t^{2}$ expresses the Coulomb repulsion for the
nearest $\sigma$-electrons and the energy dispersion is $t\left( p\right)
=3t\left( 1-\frac{p_{x}^{2}+p_{y}^{2}}{12}\right) $. Considering the explicit
form of the vertex $\Gamma_{0}$ in Eq. (10), and after summation over the
Matsubara frequencies $\omega_{n}=\left( 2n+1\right) \pi T$ one obtains 
\begin{equation}
\overset{\vee}{\Sigma}\left( p\right) =\sum \limits_{q}t\left( q\right)
\left( 1-\beta t\left( p\right) \right) \overset{\vee }{\Sigma}\left(
q\right) \frac{\tanh\sqrt{\xi^{2}\left( q\right) +\overset{\vee}{\Sigma^{2}}%
\left( q\right) }/2T}{\sqrt{\xi^{2}\left( q\right) +\overset{\vee}{\Sigma^{2}%
}\left( q\right) }}. 
\end{equation}
The search for a solution in the form $\overset{\vee}{\Sigma}\left( p\right)
=\Sigma_{0}+t\left( p\right) \Sigma_{1}$ converts Eq. (11) for the
superconducting critical temperature\ and the gap to 
\begin{equation}
1=\sum \limits_{p}t\left( p\right) \left( 1-\beta t\left( p\right) \right)
\frac {\tanh\sqrt{\xi^{2}\left( p\right) +\Sigma^{2}\left( p\right) }/2T}{%
\sqrt{\xi^{2}\left( p\right) +\Sigma^{2}\left( p\right) }}\text{,} 
\end{equation}
with the gap function $\Sigma\left( p\right) =\left[ 1-\beta t\left(
p\right) \right] \Sigma_{0}$.

Setting the gap function in Eq. $\left( 12\right) $ equal to zero, one can
derive analytically $T_{c}$ in a logarithmic approximation: 
\begin{eqnarray}
\nonumber
T_{c} & =\frac{w_{1}}{5w_{1}+4w_{2}}\sqrt{\left( w_{1}+w_{2}-r\right) \left(
w_{1}+4r\right) }\exp\left( -\frac{1}{\lambda}\right) ,  \\
\lambda & =\frac{\left( 5w_{1}+4w_{2}\right) \left( 3+5\beta w_{1}\right) }{%
\left( 2w_{1}+w_{2}+3r\right) ^{3}}\left( w_{2}-5r\right) \left[ r-\frac{%
\beta w_{1}w_{2}-2w_{1}-w_{2}}{3+5\beta w_{1}}\right] . 
\end{eqnarray}
Under the prefactor in the square root the restrictions for an energy shift $%
r$ guarantee the assumed volume of the correlated $\sigma$-band, namely $%
n_{\sigma}\geq0$ $\left( r\leq w_{1}+w_{2}\right) $ and $n_{\sigma}\leq1$ $%
\left( r\geq-w_{1}/4=-t\right) $. $T_{c}\left( r\right) $ is plotted in Fig.
4\ for different values of the parameter $\beta$, reflecting the suppression
of superconductivity with an increase of the Coulomb repulsion. 
In the MgB$_{2}$ case, $T_{c}=40$ K,
corresponds to $r=0.085$ eV\ and a dimensionless value $\beta w_{1}=8$ V$%
/3w_{1}$.
\begin{figure}
\begin{center}
\epsfig{figure=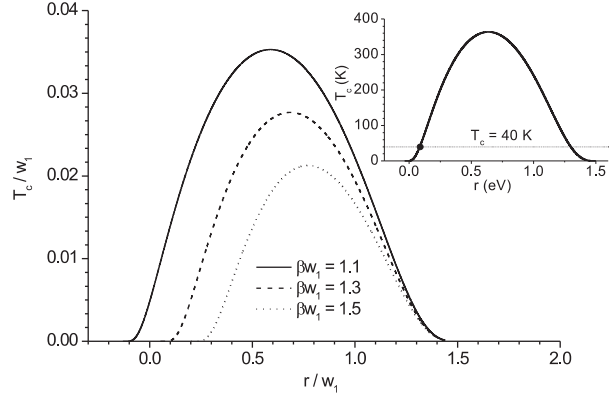,height=5.2cm}
\end{center}
\caption{The non-monotonic dependence of $T_{c}\left( r\right) $ at
different magnitudes of the Coulomb repulsion (parameter $\beta w_{1}$).
Here $w_{2}/w_{1}=8$. MgB$_{2}$ is marked in the inset with a solid circle
positioned at $r=0.085$ eV\ and $\beta w_{1}=1.2$ for $w_{1}=1$ eV and $%
w_{2}=8$ eV.}
\label{fig:nato4}
\end{figure}

The superconducting gap equation follows from Eq. $\left( 12\right) $ taken
for T=0 
\begin{equation}
1=\int \rho_{\sigma}\left( \varepsilon\right) \varepsilon\left( 1-\beta
\varepsilon\right) \frac{d\varepsilon}{\sqrt{\xi^{2}\left( \varepsilon
\right) +\Sigma_{0}^{2}\left( 1-\beta\varepsilon\right) ^{2}}}\text{.} 
\end{equation}
It defines the anisotropic superconducting order parameter of $s^{\ast}$%
-wave symmetry.

The near-cylindrical hole-like $\sigma$-Fermi surfaces in MgB$_{2}$ gives
room to calculate the superconducting DOS $\rho\left( E\right)
=\sum_{p}\delta\left( E-\sqrt{\varepsilon^{2}\left( p\right)
+\Sigma^{2}\left( p\right) }\right) $, where the gap function is 
$\Sigma=\Sigma_{0}\left( 1-\beta t\left( p\right) \right) =b\left(
1+a\cos^{2}\vartheta\right)$, 
where $a=\beta w_{1}/\left( 12\left( 1-\beta w_{1}\right) \right) $, $%
b=\Sigma_{0}\left( 1-\beta w_{1}\right) $ and $\vartheta$ is the azimuthal
angle$.$ Then the DOS normalized with respect to the normal
DOS is 
\begin{equation}
\frac{\rho\left( E\right) }{\rho_{0}\left( E\right) }=E\int\limits_{0}^{1}%
\frac{dz}{\sqrt{E^{2}-b^{2}\left( 1+az^{2}\right) ^{2}}}. 
\end{equation}

For a Coulomb repulsion such that $\beta w_{1}<1$ the parameter satisfies $%
a>0$ and the superconducting DOS becomes 
\begin{eqnarray}
\nonumber
\frac{\rho\left( b<E<\left( 1+a\right) b\right) }{\rho_{0}\left( E\right) }
& =\sqrt{\frac{E}{2ab}}K\left( q\right) , \\
\frac{\rho\left( E>\left( 1+a\right) b\right) }{\rho_{0}\left( E\right) } & =%
\sqrt{\frac{E}{2ab}}F\left( \sin^{-1}\sqrt{\frac{ab}{\left( E+\left(
1+a\right) b\right) q^{2}}};q\right) \text{,} 
\end{eqnarray}
which is expressed in terms of the complete and incomplete elliptic
integrals $K$ and $F$, respectively with modulus $q=\sqrt{(E-b)/2E}$. 

\begin{figure}
\begin{center}
\epsfig{figure=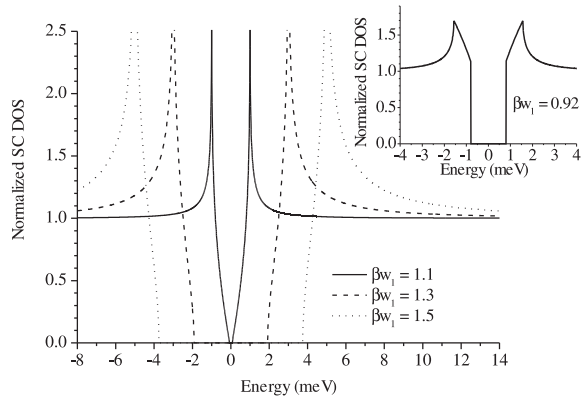,height=5.2cm}
\end{center}
\caption{The superconducting\ DOS, normalized with respect to the
normal DOS,\ for the Coulomb parameter range $\beta w_{1}>1$. At $\beta
w_{1}=0.92$, the result by Haas and Maki [15] is reproduced (inset).}
\label{fig:nato5}
\end{figure}

The
DOS\ $\left( 16\right) $ has cusps at $E$=$\pm\left( 1+a\right) b$. A
similar result was obtained in Ref. \cite{maki} 
for a non-specified parameter $%
a>0$. In our case the parameter $a$\ is controlled by the in-plane Coulomb
repulsion $V$ as the authors of Ref. \cite{4} 
noted. At $\beta w_{1}=0.92$
the DOS of Eq. $\left( 16\right) $ (see Fig. 5, inset) reproduces Fig. 1(b)
of Ref. \cite{maki}.

For our case of an enhanced Coulomb repulsion $\beta w_{1}>1$, we have
to take the parameter $a<0$ in the gap function and the
superconducting DOS (Fig. 4) is then given by 
\begin{eqnarray}
\nonumber
\frac{\rho\left( \left( 1-\left| a\right| \right) b<E<b\right) }{%
\rho_{0}\left( E\right) } & =\frac{E}{\sqrt{\left( E+b\right) \left|
a\right| b}}F\left( \sin^{-1}\frac{1}{q};\sqrt{\frac{2E}{E+b}}\right) , 
\\
\frac{\rho\left( E>b\right) }{\rho_{0}\left( E\right) } & =\sqrt{\frac
{E}{2\left| a\right| b}}F\left( \sin^{-1}q;\sqrt{\frac{E+b}{2E}}\right) . 
\end{eqnarray}
It contains
two logarithmic divergencies at $E=\pm b$ and a gap in the energy range $%
\left| E\right| <$ $\left( 1-\left| a\right| \right) b$. The two-gap ratio
is $1/\left( 1-\left| a\right| \right) $.

Measurements on MgB$_{2}$ with scanning tunneling spectroscopy \cite{9,9b}
and with high-resolution photo-emission spectroscopy \cite{10} revealed the
presence of these two gap sizes. From the ratio 3.3 between the two gaps in
Ref. \cite{9} we can extract the parameters $\left| a\right| \approx2/3$ and 
$\beta w_{1}\approx1.14$, whereas from data of
Ref. \cite{9b} one can derive $\beta w_{1}=1.21$. A value $\beta w_{1}=1.14$
can be estimated from Ref. \cite{10}. Point-contact spectroscopy \cite{10b}
shows gaps at 2.8 and 7 meV, from which we estimate a Coulomb repulsion
parameter $\beta w_{1}=1.16$. The recent study of energy gaps in
superconducting MgB$_{2}$ by specific-heat measurements revealed two gaps at
2.0 meV and 7.3 meV \cite{10c} for which $\beta w_{1}=1.13$, and a gap ratio 
$3-2.2$ \cite{10d}, for which $\beta w_{1}=1.14-1.15$. Measurements of the
specific heat of Mg$^{11}$B$_{2}$ also give evidence for a second energy gap 
\cite{10e}. Raman measurements \cite{11b} established pronounced
peaks, corresponding with gaps at 100 cm$^{-1}$ and 44 cm$^{-1}$. From these
data one can extract $a=-0.56$ and $\beta w_{1}=1.18$. At $\beta w_{1}=1$ we
have gapless like superconductivity. In this case the superconducting DOS is
linear with respect to the energy near the nodes of the superconducting
order parameter. Then the superconducting specific heat $C_{e}\sim T^{2}$
and the NMR boron\ relaxation rate $\sim T^{3}$ at low temperatures. 
From this point
of view it is interesting that the data of Ref. \cite{13} shows a $C_{e}\sim
T^{2}$ behaviour and a deviation from the exponential BCS\ behaviour in $%
T_{1}^{-1}\left( T\right) $ of $^{11}B$ \cite{14} visualized in MgB$_{2}$.

\section{Conclusions}
We have done a GL\ analysis of the superconductivity of a
two-band superconductor with particular attention to MgB$_{2}$. In that
framework, we provide the different possibilities for a T-P phase diagram,
make predictions for the role of the bands and discuss different experiments
from which crucial information can be extracted. Then we have analyzed the
superconducting properties of the material MgB$_{2}$ within the framework of
a correlated model Eq. $\left( 7\right) $. The existing electron-phonon and
non-phonon approaches to the superconducting mechanism in MgB$_{2}$ can be
separated in two groups: one pays attention to the $\sigma $-electrons and
the other to the $\pi $-electron subsystem. We have taken into account both
the correlated $\sigma $- and noncorrelated $\pi $-electrons. Analysis of
our results leads to the conclusion that superconductivity occurs in the
subsystem of $\sigma $-electrons with degenerate narrow energy bands whereas
the wide-band $\pi $-electrons play the role of a reservoir.
Superconductivity is driven by a non-phonon kinematic interaction in the $%
\sigma $-band. A lot of evidences in favour of two different superconducting
gaps can be explained by anisotropic superconductivity with an order
parameter of $s^{\ast }$-wave symmetry, induced by the in-plane Coulomb
repulsion. For an enhanced interboron Coulomb repulsion ($\beta w_{1}>1$)
the logarithmic divergencies in the superconducting DOS (Eq. $17$, Fig. 5)
are manifested by a second gap in the experiments. 
In our approach the electron-phonon coupling is hidden in the parameter $r$
. Therefore the pressure and isotope effects can be explained by the
dependence of $r\left( \omega \right) $\ on the phonon modes. From the
non-monotonic $T_{c}$ dependence on $r$\ it follows that the MgB$_{2}$\
material is in the underdoped regime (around $r\gtrsim -w_{1}/4$). For a
fictitious system A$^{2+}$B$_{2}$, where the two electrons are contributed
by atom A, the superconducting critical temperature increases with an $r$
increase. A pressure increase lowers the $\sigma $-band with an $r$ decrease
resulting in a negative pressure derivative of $T_{c}$ in agreement with
experiment \cite{20}. The bell shaped curve $T_{c}(r)$, with the MgB$_{2}$
position in the underdoped regime, shows a possibility to reach higher $T_{c}
$'s in diboride materials with an AlB$_{2}$
crystal structure. We suggest the synthesis of materials with increased $r$%
-values (e.g. with ''negative chemical pressure'') and optimized smaller
interatomic B-B distances in the honeycomb plane.

\section*{Acknowledgments}

This work was supported by the Flemish Science Foundation (FWO-Vl), the
Concerted Action program (GOA), the Inter-University Attraction Poles
research program (IUAP-IV) and the University of Antwerp (UIA).

%\section*{Figure captions}
%
%\begin{description}
%\item[Fig. 1]  A schematic view of the energy band diagram. The degenerate $%
%\sigma $-electrons are represented by a lower correlated band.%
%%
%\item[Fig. 2]  The anomalous self-energy $\overset{\vee }{\Sigma }\left(
%p\right) $ for $\sigma $-electrons. The solid line is an anomalous Green's
%function (cf. Eq. $\left( 6\right) $).
%
%\item[Fig. 3]  The non-monotonic dependence of $T_{c}\left( r\right) $ at
%different magnitudes of the Coulomb repulsion (parameter $\beta w_{1}$).
%Here $w_{2}/w_{1}=8$. MgB$_{2}$ is marked in the inset with a solid circle
%positioned at $r=0.085$ eV\ and $\beta w_{1}=1.2$ for $w_{1}=1$ eV and $%
%w_{2}=8$ eV.
%
%\item[Fig. 4]  The superconducting\ DOS, normalized with respect to the
%normal DOS,\ for the Coulomb parameter range $\beta w_{1}>1$. At $\beta
%w_{1}=0.92$, the Maki result \cite{8} is reproduced (inset) (cf. Eq. $(13)$).
%\end{description}

\end{document}